\documentclass[aps,pre,showkeys,twocolumn,showpacs,superscriptaddress]{revtex4-1}
\usepackage{graphics,graphicx}
\usepackage{dcolumn,bm}
\usepackage{mathbbol}
\usepackage{amsmath, amssymb, amsfonts,multirow}
\usepackage{color}
\usepackage{subfig}

\begin{document}

\title{Depinning transitions in elastic strings}



\author{Soumyajyoti Biswas}
\email[]{soumyajyoti.biswas@saha.ac.in}
\author{Bikas K. Chakrabarti}
\email[]{bikask.chakrabarti@saha.ac.in}

\affiliation{
Theoretical Condensed Matter Physics Division, Saha Institute of Nuclear 
Physics, 1/AF Bidhannagar, Kolkata-700064, India.\\
}

\date{\today}

\begin{abstract}
\noindent We study the depinning transitions of elastic strings in disordered media in two different cases.
We consider the elastic forces to be of infinite range in one case, where the magnitude is proportional to
the extension of the string. The critical points and elastic behavior can be estimated to some extent in this
case. The exponents are estimated numerically and scaling relations are found to be obeyed. 
We have also considered a model where the elastic force is constant in magnitude. The critical points can again be argued
and the critical exponents are numerically estimated. The scaling relations are well obeyed. Both of these 
models do not fall into known universality classes.          
\end{abstract}

\pacs{}

\maketitle

\section{Introduction}
\noindent Dynamics of elastic strings, driven through a medium with quenched disorder \cite{book}, has served for modelling of the dynamics of 
diverse physical phenomena such as 
charge density waves \cite{Gruner}, vortices in type-II superconductors \cite{vort}, magnetic domain walls in external magnetic field \cite{domain}
and also propagating
fracture fronts \cite{prop1} (for a recent review see \cite{bouchaud}).

When an elastic string is driven through a random medium, the forces that act on the string are the following: the uniform external force ($f$),
the random pinning force due to impurities and the local or non-local force due to elastic stretching of the string ($f_{el}$).
When the external driving force is small enough, the string as a whole would not propagate (in the long time limit). In that case
the system is said to be pinned by the impurities present in the medium. On the other hand, when the force exceeds a threshold $f_c$,
the string will propagate through the disordered medium with a finite velocity (in the long time). At $f=f_c$, string is just depinned.

Global failure processes due to fracture, has been studied using the tools of statistical mechanics before \cite{rmp}. 
However, a long studied approach to study depinning transition problems have been to deal the depinning of the string as a dynamical phase transition.
Much attention have been paid in this area 
both theoretically as well as experimentally (see e.g., \cite{fis,zapperi,he,kim,maloy1,maloy2}).
Since in the depinned state the average velocity $v$ of the string is non-zero and in the pinned phase it is zero, the velocity is taken 
as the order parameter for this transition. As it is a second order phase transition, the order parameter should scale as
$v\sim(f-f_c)^{\theta}$ where $\theta$ is velocity or order parameter exponent. Other critical exponents and scaling relations
can also be derived to formally classify a universality class of this transition.

The most studied universality classes in the depinning transitions are the Edwards-Wilkinson (EW) \cite{EW,EW1,EW2} and Kardar-Parisi-Zhang (KPZ) \cite{KPZ,KPZ1,KPZ2} 
universality classes (see Ref.\cite{book} for details). Apart from these models where the elastic force is essentially local (depends on the positions of 
the nearest neighbors of an
element), there are some non-local extensions, which are relevant for fracture front propagation, where the elastic force on an
element has contributions from all other elements. However, the amount of contribution decreases in a power law (mostly inverse
square) with the distance from the said element (see e.g., \cite{krauth}). A mean field model, in which the elastic force on an
element depends on its distance from the average height, has also been studied \cite{mf}.

In this paper we study two different limits of the form of the elastic force. First we study the case where the elastic force
is proportional to the total elongation of the string and has equal magnitude at each point on the string. The sign, however,
depends upon the sign of the local curvature. Clearly, in this case, the elastic force is of infinite range. Using this fact, we 
argue to find the critical point at which the string is just depinned. The critical behavior is obtained by integrating
the equation of motion. The exponent values satisfy usual scaling relations. 

Secondly, we consider a case, where the magnitude of the elastic force is a constant and the sign depends upon the local curvature.
Furthermore, the pinning forces remain constant for a given point on the string. This is in contrast with the convention that
the pinning force changes its value after certain intervals along the propagation path of the string.  

The paper is organised as follows: In the next section we describe the model and the numerical methods used in obtaining 
the height profile. Next we present the scaling analysis and estimate the critical exponents and the scaling laws satisfied
by them. Finally we discuss these results and conclude.

 \section{Models and methods}
\noindent As mentioned before, an elastic chain driven through a medium with quenched disorder faces three types of forces: External driving
force ($f$), elastic force ($f_{el}$), which tends to conserve the length of the string and a random pinning force ($f_p$) due to quenched disorder.
The (discretized) equation for the time evolution of the ``height'' profile of the string is
\begin{eqnarray}
h_i(t+1)=h_i(t)+f_{el}+f_p+f
\end{eqnarray} 
where, $h_i(t)$ denotes the height of the $i$-th element of the string at time $t$. 
The movement of the string is allowed only in the forward direction.

\begin{figure}[tbh]
\centering \includegraphics[width=1.0\linewidth]{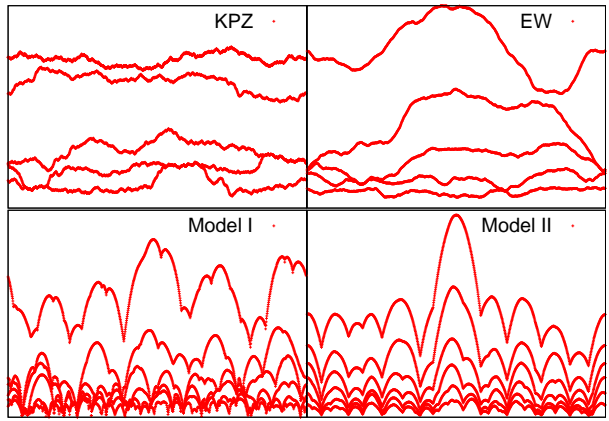}
   \caption{The height profile $h_i(t)$ is plotted against $i$ for different time instants for EW, KPZ and the two models studied here. The time
index increases upwards.}
\label{opd0}
\end{figure}

One can define an average velocity (at time $t$) of the string as
\begin{equation}
v(t)=\frac{1}{N}\sum\limits_{i=1}^N(h_i(t)-h_i(t-1)).
\end{equation}
The `pinned' and `depinned' phases of the string is determined from the fact whether the above quantity vanishes
or attains a finite value in the long time limit. This is analogous to non-equilibrium active-absorbing type phase transitions
 with velocity ($v(t)$) playing the role of the order parameter. The external force $f$ is usually taken as the `ordering field'.
Therefore, near the transition point one would expect 
\begin{equation}
v(t\to \infty) \sim (f-f_c)^{\theta},
\end{equation}    
where, $f_c$ is the critical value of the external force where the chain is just depinned. Other critical exponents 
and scaling relations (discussed later) can also be obtained, which define the universality class of the transition.
The universality classes are dependent upon what form of the elastic force is chosen. For example, when the elastic force
is simply proportional to the local curvature ($\nabla^2 h$), the universality class is Edwards-Wilkinson (EW); when 
it is proportional to the extension of the chain in the nearest neighbor, it is KPZ type and so on. All these cases are for
uncorrelated noise, the cases of correlated noise change the situation a lot (see \cite{book}) which we shall not discuss here.

Below we study two models, where this elastic forces are different. In one model, the elastic force is infinite range
and its magnitude is proportional to the total extension of the string and the sign of the force depends upon the local
curvature. In another model, the magnitude is taken as a constant and the sign still depends on the sign of the local curvature
(Fig.~\ref{opd0} depicts height profiles of these two and EW and KPZ models at critical point for different time instances).

\subsection{Model I}
\noindent Here we consider an elastic chain of infinite range interaction driven through a medium with quenched disorder. 
We assume linear elastic behavior, such that the net extension of the chain generates an elastic force which is felt
by all element of the chain. In particular, we consider the following equation of motion for the height profile

\begin{eqnarray}
\label{model}
h_i(t+1) &=& h_i(t)+f_{el}+\eta_i(h_i(t))+f \nonumber \\
&=& h_i(t)+\frac{1}{L}\sum\limits_j\left[\sqrt{(h_j(t)-h_{j+1}(t))^2+1}-1 \right] \hat{C_i} \nonumber \\
&& +\eta_i\left(h_i(t)\right)+f
\end{eqnarray}
with $|f_{el}|=\sum\limits_j\left[\sqrt{(h_j(t)-h_{j+1}(t))^2+1}-1 \right]/L$, $\hat{C_i}=sgn(h_{i+1}+h_{i-1}-2h_i)$ and $h_i(t)$ denotes the height of the $i$-th element of 
the discretized chain at time $t$, $\eta_i$ denotes  the random uncorrelated
quenched noise in the pinning force (the nature of which we further illustrate below), $f$ denotes the constant external force applied  
uniformly (on each element of the chain)
and $\hat{C_i}$ denotes the sign of the local curvature i.e., the net extension force acts to restore the flat initial 
condition for the chain thereby creating a restoring force for the locally advanced elements and pushes the locally
behind elements.

The nature of the disorder in pinning force $\eta_i$ is such that with probability $p$ it is zero and with probability $1-p$ it is uniform between 0 and $-a$. 
$p$ acts here like a probability of presence of vacancies within the material (in the following we will study $p$ values ranging from $0$ to $0.5$). 
To take the pinning strength of the material to have uniform values is a matter of convention, 
the critical behavior does not change if one takes it differently (triangular, say). 
It is to be noted here is that, unlike general convention, the disorder $\eta_i$ here has no positive value. In our simulations, we took an
elastic chain of length $L$ ($=5000$ mostly), with periodic boundary conditions.

\subsubsection{Integration of the equation of motion}
\noindent The critical properties (exponent values) of 
this transition are to be 
found out by numerically
integrating the equation of motion (see e.g., \cite{kim}). This is done following the equation:

\begin{eqnarray}
\label{inte}
h_i(t+\Delta t) && =h_i(t)+ G_i(t) \Delta t \qquad \mbox{if} \qquad G_i(t)> 0 \nonumber \\
&& =h_i(t) \qquad \mbox{otherwise}
\end{eqnarray}

\noindent where $G_i(t)=f_{el}\hat{C}_i+\eta_i\widetilde{\left(h_i(t)\right)}+f$
and $\widetilde{h_i(t)}$ denotes the integral part of $h_i(t)$, $\Delta t$ is a small interval of time, mostly taken as $0.01$ but
our results do not change even if it was made smaller (say, $0.005$). Clearly, it is ensured in the dynamics that any element never takes 
backward step, which is unphysical from the point of view of fracture front propagation. 

In the following, we have presented numerical results for $L=5000$, $\delta t=0.01$, $p=0.5$ and averaging over $100$ configurations. It is checked that
making $\delta t=0.005$ or $L=10^4$ does not change the conclusions.

\subsection{Model II}
\noindent In this case we make two major changes. First, the elastic force is taken as a constant and its magnitude is determined by the
local curvature. In this sense, it is a model with short range interaction. Secondly, we keep the quenched disorder for a given site
fixed in height, i.e., $\eta_i$ is no longer a function of (integral part of) $h_i(t)$. The evolution equation for height, therefore, becomes
\begin{equation}
h_i(t+1)=h_i(t)+F\hat{C}+\eta_i+f
\label{flock}
\end{equation} 
where $F$ is the constant magnitude of the elastic force and other symbols have same meanings as before. 

One can apply this model for `depinning' transition in a flock of birds flying in a line (see e.g., \cite{flock1} for references related to bird flocking). 
The elastic force denotes the tendency of
a bird to fly along with its two nearest neighbors, the pinning force may be considered as some measure of `(un)fitness' of a bird (that is why it is
taken as independent of height or in this case, distance travelled), 
and the external force is the urge or necessity of flying (lack of food, presence of enemy etc.). Of course, $v\to 0$
limit does not apply for a flock of flying birds. But here we can measure the velocity from the rest frame of the slowest moving bird.
Then the `depinning' transition would indicate whether the birds fly `together' or gets scattered in the long time limit. If the average velocity
of the flock is zero with respect to the slowest moving bird, then they fly `together' or the flock is `pinned'. If, however, the average velocity is finite
in the rest frame of the slowest bird, then in the long time the flock will be `scattered' or the flock is said to be `depinned'.  

The nature of disorder used has similar properties as before. It has a random value between $[0:-1]$ with probability $1-p$ and zero otherwise. Below 
we present results for $p=0$ case, but the exponent values do not change for finite $p$ values.

The evolution rule (Eq.(~\ref{flock})) is analogous to the equation of motion presented above for an elastic chain. But the major difference is that
in this case it is not an `equation of motion' but only a rule chosen for the movement of the birds. Therefore the choice of $\Delta t$ is arbitrary
and we fix it to unity in this case. 

\subsection{Scaling analysis}
\label{scal}
\noindent Using the numerical method described above we measure the height profile of the chain with time. At and near the 
depinning transition point ($f=f_c$), several quantities related to the height profile show interesting behavior. To begin with,
we measure the average of the velocities of the elements, which is the order parameter $v(t)=d\langle h_i(t)\rangle_i/dt$. Above 
the depinning point, the average velocity initially decreases, but at long time saturates to a finite non-zero value depending
on the external force. On the other hand if the force is below the depinning point, the average velocity decreases faster than a power law
and eventually becomes zero. Just at the critical point the velocity decreases as a power law
\begin{equation}
v(t)\sim t^{-\delta}.
\label{eq:delta}
\end{equation} 
Not only does this study give the value of the exponent $\delta$, it is also useful in locating the critical point. 

Considering the long term saturation values above the depinning transition, one would expect the growth
\begin{equation}
v(t\to \infty,f)\sim (f-f_c)^{\theta},
\label{eq:theta}
\end{equation}
where $\theta$ is the velocity exponent. 

Another interesting quantity is the mean square fluctuation or the width of the surface, which is defined as
\begin{equation}
W(L,t)=\langle \frac{1}{L}\sum\limits_i\left[ h_i(t)-\langle h(t)\rangle\right]^2 \rangle^{1/2}.
\label{eq:width}
\end{equation}
This will show a growth of the form $W(t)\sim t^{\beta}$.

Also it is argued that at criticality, the velocity of the chain is essentially governed by avalanches and an avalanche
takes a time $t$ to advance a distance equal to its width \cite{krauth}. 
Therefore, one would expect $v(t)\sim W(t)/t$, which would imply $\delta+\beta=1$.

Finally, the off-critical scaling of the average velocity has the form
\begin{equation}
v(f,t)\sim t^{-\delta}F\left(t|f-f_c|^{\nu}\right),
\label{eq:nu}
\end{equation}
where $\nu$ is the exponent for the diverging correlation time. It is expected to satisfy $\nu=\theta/\delta$.

Below we present numerical results for the two models estimating the above mentioned critical exponent values and show that the scaling relations
are obeyed.
\section{Results}
\noindent The critical points and for that matter the phase diagrams for both the models can be argued to some extent. However, the
critical behavior, i.e., the values of the exponents are to be found out numerically as indicated before. In this section we present the 
phase diagrams and critical behaviors for both of the models. 

\subsection{Model I}

\subsubsection{Elastic behavior and critical points}

\begin{figure}[tbh]
\begin{center}
 \includegraphics[width=1.0\linewidth]{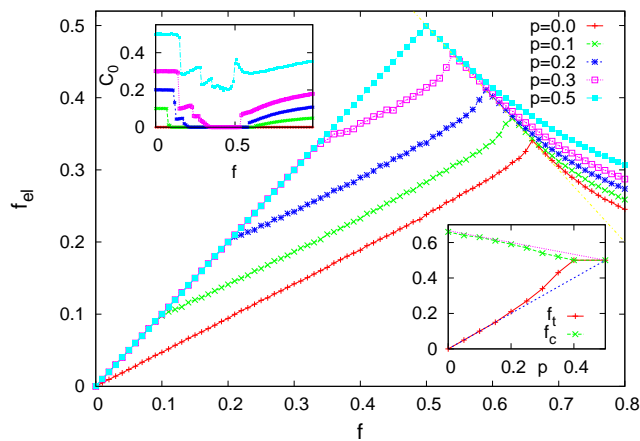}
\end{center}
   \caption{Model I: The elastic force $|f_{el}|$  is plotted against 
external force for different values of $p$. It is seen that the response is piecewise linear, with a change
of slope associated with elimination of zero pinning force on the string. Top left inset confirms this statement, it also shows a increase in density of
zero pinning force ($C_0$) after depinning. The bottom right inset shows the variation of the point where change of slope occurs ($f_t$) and also the 
depinning force $f_c$). The continuous lines are predictions for small $p$ values (see text). For the simulation data shown, we have used $L=5000$ and
Euclidean metric for measuring the length of the string.}
\label{elastic}
\end{figure}

\noindent Due to the infinite nature of the elastic force, it is possible to predict the critical point and to some extent the elastic behavior
of the chain in the pinned state. Fig.~\ref{elastic} depicts the variation of elastic force ($f_{el}$) with applied external force ($f$) for different $p$ values.
Now, when the string is just depinned, the element facing the maximum pinning force $f_p^{max} (=1)$ must be depinned. So, the condition for depinning is
\begin{equation}
f+f_{el}=f_p^{max}=1.
\label{condition1}
\end{equation}
From Fig.~\ref{elastic} it is seen that for all $p$ values the elastic force increases with external force upto the point when their sum become unity. 
At that point the depinning transition occurs. As we will elaborate later, independent estimate of the critical points agree well with this measure.
Now to understand the elastic behavior, first note that when the string is pinned, the points facing the weakest pinning must also be pinned. Of course,
the weakest pinning for a particular pinned configuration of the string need not be the weakest pinning possible (i.e., 0). It will actually depend 
upon the external force. The pinning condition is
\begin{equation}
f_p^{min}+f_{el}=f.
\label{condition2}
\end{equation}     
Before this condition is reached, both terms in the left hand side would increase in magnitude until the condition is satisfied. This can be understood from the 
observation that the sites facing the minimum pinning would move forward and eventually face a stronger pinning, thereby increasing the first term. The overall
elongation of the chain during this process would increase the second term. The minimum value of the both terms for which the above condition is satisfied
will be when $f_p^{min}=f_{el}=\frac{1}{2}f$. This is also verified numerically. This indicate the slope of the $p=0$ line to be $0.5$, which is indeed the case 
(see Fig.~\ref{elastic}).

For the cases with finite $p$, $f_p^{min}=0$ until the string attains a configuration where there is no longer a zero pinning force in its surface.
Until that point, however, Eq.~(\ref{condition2}) implies $f_{el}=f$, which is actually the case seen in Fig.~\ref{elastic}. Once there is no more zero pinning on the string,
it behaves as $p=0$ case and thereby attains a slope $0.5$ as before. It is also checked numerically that the change of slope is indeed associated with
the absence of zero pinning in the string (see top left inset of Fig.~\ref{elastic}). Furthermore, when $p$ is very small, a small value of $f$ 
is required to move all the sites having zero pinning
and to reach the elimination of zero pinning point.
In the first approximation one can assume that essentially the zero pinned sites are the only ones to move as $f_{el}$ will be small as well. Under this approximation,
the change of slope would occur at $f=p$. As one can see (bottom right inset of Fig.~\ref{elastic}), this is valid almost upto $p=0.2$. 
In this regime, the critical point would be  $f_c=(2-p)/3$, which is
indeed the case for small $p$. For large values of $p$, there cannot be a pinned configuration where there is no zero pinned site. Therefore, for large values
of $p$, the $f_{el}=f$ relation is always valid, as can also be seen numerically.

\subsubsection{Critical behavior and scaling relations} 
\noindent Here we present the numerical results of the scaling analysis mentioned in \ref{scal} for Model I.

In Fig.~\ref{unscaled} 
the average velocities are plotted for different values of the external force on both sides of the estimated critical point
$f_c=0.500\pm 0.001$, which matches very well with the estimate $a/2$ argued above (for $p\ge 0.5$). In this log-log plot, the power-law
decay of the velocity at the depinning point comes as s straight line from which $\delta$ is estimated to be $0.60\pm0.01$.
 
\begin{figure}[tbh]
\begin{center}
 \includegraphics[width=1.1\linewidth]{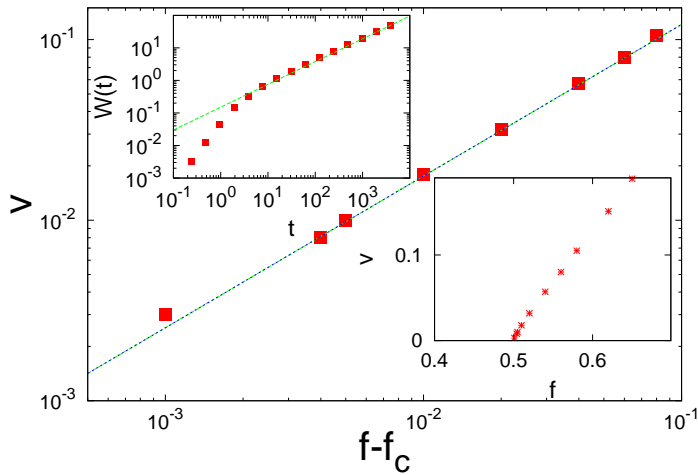}
\end{center}
   \caption{Model I: The saturation values of the average velocities (order parameter) are plotted for different external 
force $f=0.5005$, $0.501$, $0.502$, $0.504$, $0.506$, $0.508$. 
The log-log plot gives the estimate for $\theta\approx0.83\pm0.01$. System size is $L=5000$. The line is guide to the eye. The bottom right inset shows the variation 
of order parameter with external force. The top right inset shows the log-log plot of $W^2(t)$ with $t$ at $f=f_c$ giving $2\beta=0.70\pm0.05$. Euclidean metric is used
for measuring the length of the string.}
\label{theta}
\end{figure}

Then inn Fig.~\ref{theta} we plot the saturation values of velocity in the depinned region.
Knowing the critical point accurately from the above analysis, the so called velocity exponent $\theta$ is estimated to be $0.83\pm 0.01$.

\begin{figure}[tbh]
\begin{center}
 \includegraphics[width=1.0\linewidth]{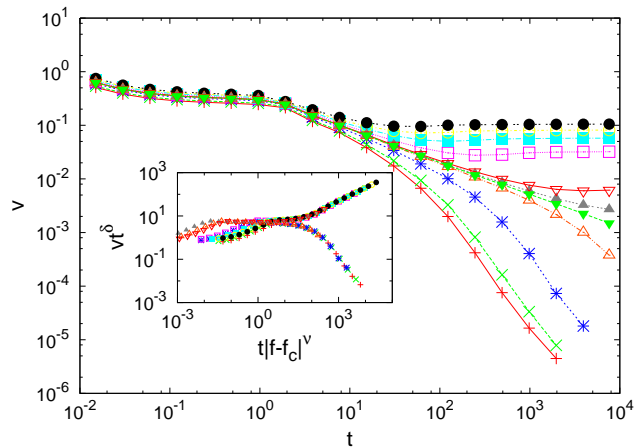}
\end{center}
   \caption{Model I: The average velocities are plotted for different values of external forces above, below and at criticality. In the inset
 $t|f-f_c|^{\nu}$ is plotted against $vt^{\delta}$ for different values of $f$. Estimating $\delta$ ($\approx 0.60\pm0.01$), the fitting value of
 $\nu$ is found to be $1.35\pm0.05$. The system size is $L=5000$ and $\Delta t =0.01$ for all the simulations. Euclidean metric is used
for measuring the length of the string. }
\label{unscaled}
\end{figure}

In top left inset of Fig.~\ref{theta} the growth of $W^2(t)$ at the critical point is plotted against time. The power law fit gives $2\beta=0.70\pm0.05$.
This growth, however, has a bound as in the case of other mean-field models \cite{mf}. 
From our estimate $\delta+\beta=0.95$, which is in reasonable agreement with the scaling relation.

In Fig.~\ref{unscaled} we plot $vt^{-\delta}$ against $t|f-f_c|^{\nu}$. We already know the value of $\delta$. So by only
tuning $\nu$ we get a data collapse and the corresponding estimate for $\nu=1.35\pm 0.05$. From our previous estimates of
$\theta$ and $\delta$ we find $\nu=\theta/\delta\approx 1.38$, which is in reasonable agreement with our independent estimate.

\subsubsection{Universality}
\label{univ}
\noindent We have checked the universality of the above results by changing the metric of measuring the distance or the
elongation of the string. Instead of measuring the length using the Euclidean metric as mentioned above, we have used 
the rectilinear length or the Manhattan length of the string. In general, the rectilinear or Manhattan distance between
two points $(x_1,y_1)$ and $(x_2,y_2)$ is defined as $|x_1-x_2|+|y_1-y_2|$. Using this metric, Eq.~(\ref{model}) will get modified to
\begin{eqnarray}
\label{model1}
h_i(t+1)= && h_i(t)+\frac{1}{L}\sum\limits_j\left[|h_j(t)-h_{j+1}(t)| \right] \hat{C_i} \nonumber \\
&& +\eta_i\left(h_i(t)\right)+f
\end{eqnarray}
with symbols having usual meanings. 

\begin{figure}[tbh]
\begin{center}
 \includegraphics[width=1.0\linewidth]{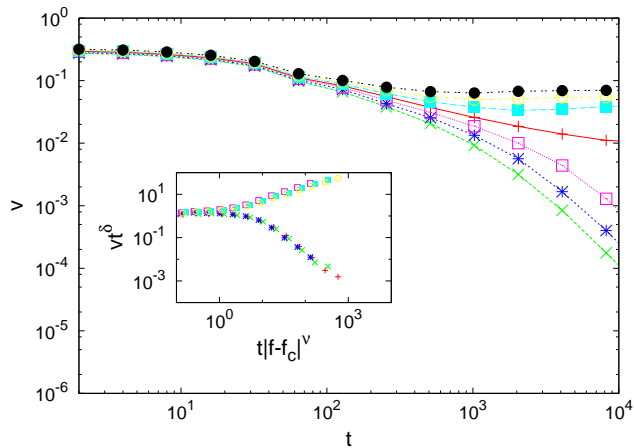}
\end{center}
   \caption{Model I: The average velocities are plotted for different values of external forces above, below and at criticality for Manhattan distance metric. In the inset
 $t|f-f_c|^{\nu}$ is plotted against $vt^{\delta}$ for different values of $f$. Estimating $\delta$ ($\approx 0.60\pm0.01$), the fitting value of
 $\nu$ is found to be $1.35\pm0.05$. The system size is $L=5000$ and $\Delta t =0.01$ for all the simulations.}
\label{unscaled}
\end{figure}

Note that the force due to extension is still felt equally by each element of the string. Therefore, the estimate of the
critical point, as mentioned before, remains unchanged (i.e., half of the maximum pinning force). More importantly, the
values of the exponents ($\delta$, $\theta$, $\nu$ and $\beta$) remains same upto the error bar of their estimates 
($\delta=0.60\pm0.01$, $\theta=0.84\pm0.01$, $\nu=1.35\pm0.05$, $\beta=0.35\pm 0.05$; satisfying very closely the relations $\delta+\beta=1$ and $\nu=\theta/\delta$).
\subsection{Model II} 
\subsubsection{Phase diagram}

\begin{figure}[tbh]
\begin{center}
 \includegraphics[width=1.0\linewidth]{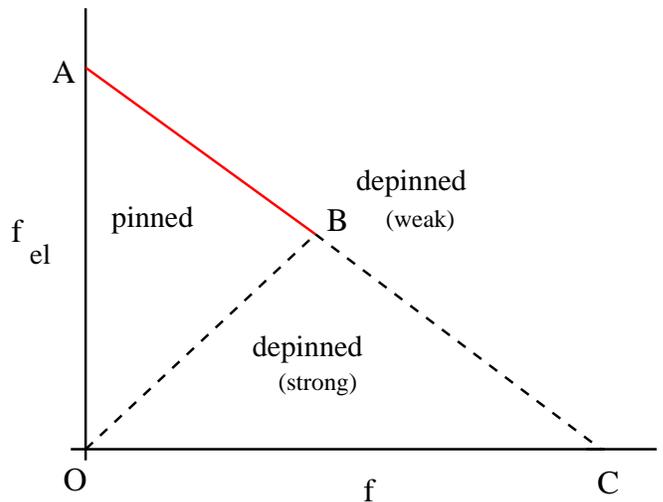}
\end{center}
   \caption{The phase diagram for Model II depicting pinned phase and two types of depinned phase. The line ABC follows the equation
$f_{el}+f=f_p^{max}$, the line OB is $f_{el}=f$. The strongly depinned phase is
where Eq.~(\ref{strong}) is satisfied. Here each site moves forward independently. Therefore ballistic motion is seen giving
linear increase in the width of the profile. The boundary between pinned and strongly depinned states does not represent 
a phase transition as in the later phase no cooperative phenomenon takes place. The weakly depinned region is characterised by sublinear growth of
the width of the profile. The transition between pinned and weakly depinned states is a phase transition whose exponents are reported (see text).}
\label{ph-dia-i}
\end{figure}

\noindent Consider two sites $i$ and $j$. Regarding the pinning force, let $i$-th site faces a force which is a local minimum (in magnitude) and $j$-th site 
is the nearest site having pinning force at a local minimum. Initially, $j$-th site will move forward and $i$-th site will 
lag behind. Therefore, the forces on $i$-th and $j$-th site will 
respectively be
\begin{eqnarray}
F^i &=& f_{el}-f_p^i+f \nonumber \\
F^j &=& -f_{el}-f_p^j+f,
\end{eqnarray}
where $f_p^a$ denotes pinning force at $a$-th site. The above equation indicates that it is only possible for the $i$-th site to come in the same level as
that of $j$-th site is when
\begin{equation}
2f_el > \Delta f_p,
\end{equation} 
where $\Delta f_p$ denotes that difference between the pinning forces at the two sites. Again, the $i$-th site can move forward only when
\begin{equation}
f_p^i\le f_{el}+f.
\end{equation}
The above two conditions together imply
\begin{equation}
f_{el}\ge f.
\label{strong}
\end{equation}
As we shall see later, this condition prevents independent ballistic motion of the birds; this is also clear from the fact that $f_{el}$ is
a kind of nearest neighbor interaction that invokes cooperitivity and the width of the profile increases sublinearly with time in the depinned state
(we call this weakly depinned state; see Fig.~\ref{ph-dia-i}).

Note further that the maximum pinning force a site may face is $f_p^{max}$. Therefore, the condition for depinning would be
\begin{equation}
f_p^{max}\le f_{el}+f,
\end{equation}
with the equality determining the critical points. 

Now consider the case when $f>f_{el}$. Irrespective of the magnitude of $f$, this is a depinned state. This is because, as the lower limit of the 
pinning force is zero, for any value of $f$, some sites, which have pinning less than $f$, will be moving. Now, as $f_{el}<f$, those sites will never stop.
This is, therefore, a trivial case of depinning where the particles move essentially independently and the `width' of the profile
increases linearly with time; we called this strongly depinned state (see Fig.~\ref{ph-dia-i}).
\subsubsection{Critical behavior and scaling relations}
\begin{figure}[tbh]
\begin{center}
 \includegraphics[width=1.0\linewidth]{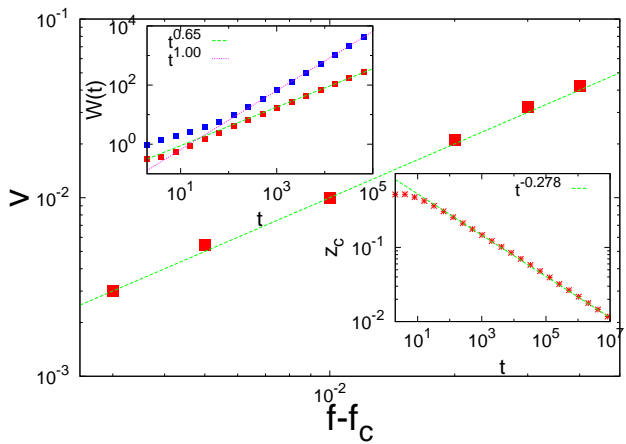}
\end{center}
   \caption{Model II: The top left inset shows the growth of the width for strongly (upper curve) and weakly (lower curve) depinned states. 
In the strongly depinned state, the width increases linearly and in the weakly depinned state it increases sublinearly (with
an exponent $\beta=0.65\pm0.01$). The main figure shows variation of the order parameter in the pinned and weakly depinned transition. It gives $\theta=1.00\pm0.01$
as the order parameter exponent. The bottom right inset shows the decay of the density of sites ($z_c$) having positive curvature at the pinned and weakly depinned
transition point. This gives a power-law decay with exponent $0.278\pm0.005$. This feature is unique for Model II. For simulations system size was $L=10^4$. The 
data presented here is for the critical point $f_c=0.4, f_{el}^c=0.6, f_p^{max}=1$. For other points along the phase boundary (between pinned and weakly depinned state)
the exponents are same.}
\label{betatheta-i}
\end{figure}

\noindent Here we report the numerical estimates of the exponent values for the pinned and weakly depinned transition. 
In Fig.~{\ref{betatheta-i}}, the order parameter (measured in the rest frame of the slowest moving site) is plotted against $f-f_c$ 
to estimate the order parameter exponent. It is found to
be $\theta=1.00\pm 0.01$. The width of the profile increases in a power law with exponent $\beta=0.65\pm0.01$. It is also shown that
in the strongly depinned region, the width increases linearly. Another quantity is measured here, which is the concentration of the sites
having positive (open upwards) curvature ($z_c$). This quantity shows non-trivial power-law relaxation only for this model with
an exponent $0.278\pm0.001$. 

\begin{figure}[tbh]
\begin{center}
 \includegraphics[width=1.0\linewidth]{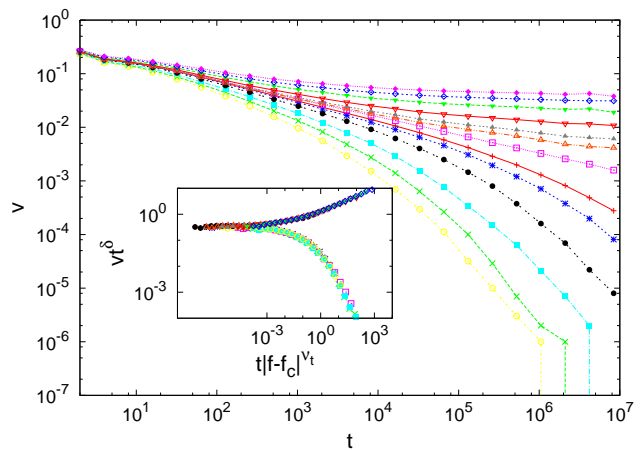}
\end{center}
   \caption{Model II: The time dependence of the order parameter (velocity) is plotted for different values of $f$ around $f_c$ (at the same critical point 
 mentioned in the previous figure) measured in the rest frame of the slowest moving site.
At $f_c$ the velocity decays with an exponent $\delta=0.34\pm 0.01$. The inset shows the data collapse to find the exponent $\nu=2.95\pm0.05$. The simulations
are for $L=10^4$ and the estimates of the exponent values remains same along the pinned-weakly depinned phase boundary.}
\label{nui}
\end{figure}

In Fig.~{\ref{nui}}, the relaxation of the order parameter in time is shown near a critical point. At the critical point, the velocity
decays in a power law with an exponent $\delta \approx 0.34\pm0.01$. Also, using the scaling techniques discussed in \ref{scal}, the exponent $\nu$
is estimated to have the value $2.95\pm0.05$. 

As one can see from the above discussions, the scaling relations $\delta+\theta=1$ and $\delta=\theta/\nu$ are almost satisfied. 

\begin{table*}
\caption{The exponents values for other known universality classes are compared with the ones presented here.}
\begin{center}
    \begin{tabular}{ | l | l | l | l | p{2cm} |}
    \hline
    Models & $\beta$ & $\theta$ & $\nu$ & $\delta$ \\ \hline
    EW \cite{amaral} & $0.85\pm 0.03$ & $0.24\pm0.03$ & $1.73\pm0.04$ & ~~~~~~ - \\ \hline
    KPZ \cite{amaral} & $0.67\pm0.05$ & $0.64\pm0.12$ & $1.35\pm0.04$ & ~~~~~~ -\\ \hline
    $1/r^2$ \cite{krauth} & $0.495\pm 0.005$ & $0.625\pm 0.005$ & $1.625 \pm 0.005$  & ~~~~~~ - \\ \hline
    MH \cite{kim} & $0.841\pm0.005$ & $0.289\pm0.008$ & $1.81\pm 0.1$ & $0.160 \pm 0.005$ \\ \hline
    Model I & $0.35\pm0.05$ & $0.83\pm0.01$ & $1.35\pm0.05$ & $0.60\pm0.01$ \\ \hline
    Model II & $0.65\pm0.05$ & $1.00\pm 0.01$ & $2.95\pm0.05$ & $0.34\pm0.01$ \\ \hline 
    \end{tabular}
\end{center}
\end{table*}

\section{Summary and Discussions}
\noindent In the present work, we study numerically the scaling properties of the dynamics of  elastic string driven through a disordered medium.
We have considered two models here, in one case the elastic force is of infinite range and in second it is short range and constant.

By infinite range we mean that the elastic
force in an element depends upon the entire elongation of the chain (see Eq.~(\ref{model})). Of course, the sign (direction) of the force is not uniform for all 
the elements; it depends upon the sign of the local curvature. We have employed both Euclidean metric and Manhattan metric for determining the string length. 
Apart from the elastic force, we have assumed a random pinning force 
and an external force which is applied to depin the chain. 
Using the non-locality of the elastic force, we have argued that the critical force at which the chain will be depinned is
$f_c=(2a-p)/3$ for small values of $p$, where $a$ is the maximum pinning force and $f_c=a/2$ for $p\ge 0.5$. 
 By knowing the critical force (for $p\ge 0.5$), we integrate the equation of motion for the height profile (see Eq.(\ref{inte})) and perform standard
scaling analysis to get the critical exponents. In particular, we have found the power-law decay of the average velocity at the 
critical point (giving $\delta$), saturation values of the velocities in the depinned region gives $\theta$. From off-critical scaling the exponent for the 
diverging temporal correlation length ($\nu$) is found. The rms width grows in a power-law at the critical point giving the exponent $\beta$. 
All these exponent values fit well with  the scaling relations  $\delta+\beta=1$ and 
$\nu=\theta/\delta$. The estimates remain consistent within error bars when the same is done for other values of $p$ as well as by changing the metric of
the distance measurement. As discussed already, $\beta$, $\theta$, $\nu$ and $\delta$ values are $0.35\pm0.05$, $0.83\pm0.01$, $1.35\pm0.05$ and $0.60\pm0.01$
respectively for Euclidean metric (see \ref{scal}) and $0.35\pm0.05$, $0.84\pm0.01$, $1.35\pm0.05$ and $0.60\pm0.01$ respectively for Manhattan metric (see \ref{univ}).
These are summarised in Table 1 and are also compared with the other universality classes.
Note that the estimated exponent value of $\theta \approx 0.83\pm0.01$ in our model is very close to the experimental value
of $\theta \approx 0.8\pm0.1$ by  Ponson \cite{ponson, bouchaud}.

We have also studied a model (as mentioned before) where the elastic force is of constant magnitude for each element, although its sign (as before)
depends upon the sign of the local curvature. In this model we have taken the pinning force as constant for each element during its entire motion. 
This model is relevant for the study of flying flock of birds. As one can see, the pinned force may then represent the (un)fitness of the birds and
will remain more or less constant. The `elastic force' in this case does not depend upon 
the entire chain of birds, but is only of constant magnitude with the sign being determined 
by the nearest neighbors (local curvature). As one can see, the zero velocity for a flying flock of birds has no meaning. But here we consider the velocity
of the flock from the rest frame of the slowest moving bird. The `depinning' transition in this case implies that in the `depinned' phase,
 even from the rest frame of the slowest moving bird,
the `width' of the flock increases with time as opposed to its remaining constant in the `pinned' phase. As one can see from Fig.~{\ref{ph-dia-i}},
three phases are present. In the pinned phase, all birds finally have the velocity of the slowest moving one, thereby keeping the width of
the profile constant. In the strongly depinned case, these is no sense of cooperitivity (see Eq.~(\ref{strong})) and the birds move independently,
making the width increase linearly (see Fig.~{\ref{betatheta-i}}). In the weakly depinned phase, however, the width increases sublinearly. The other 
exponents of the this transition (pinned to weakly depinned) is summarised in Table 1 
and also are compared with other universality classes. As one can see, the scaling relations $\delta+\beta=1$
and $\delta=\theta/\nu$ are satisfied within the error bars of the estimates.

In summary, we have studied  two models of depinning transitions for `elastic' strings driven through disordered (quenched) media. The exponents
estimated do not fall into known universality classes. As argued before the models (I \& II) seem to be relevant for fracture 
front propagation and flocking of birds respectively.

\end{document}